\documentclass[aip,amsmath,amssymb,nofootinbib,numerical,reprint]{revtex4-1} 

\usepackage[american]{babel}
\usepackage{dcolumn}
\usepackage{bm}
\usepackage{graphicx}	
\usepackage{float,wrapfig}
\usepackage[utf8]{inputenc}
\usepackage[T1]{fontenc}
\usepackage{longtable,tabularx}
\usepackage{multirow}
\usepackage{booktabs}
\usepackage[space]{grffile} 
\usepackage{url,hyperref,lineno,microtype}
\usepackage[nameinlink,noabbrev]{cleveref}
\usepackage{wasysym} 
\usepackage[caption=false]{subfig}	
\usepackage{xcolor,soul}
\usepackage{physics}
\usepackage{lineno}
\hypersetup{
    colorlinks,
    linkcolor={gray!30!cyan},
    citecolor={gray!30!cyan},
    urlcolor={gray!30!cyan}
}

\let\oldsim\sim 
\renewcommand{\sim}{{\oldsim}}
\newcommand{\comment}[1]{}
\newcommand*\diff{\mathop{}\!\mathrm{d}}

\newcommand*{\figureref}[2][]{%
  \hyperref[{fig:#2}]{%
    Figure~\ref*{fig:#2}%
    \ifx\\#1\\%
    \else
      ~(#1)%
    \fi
  }%
}

\newcommand*{\figref}[2][]{%
  \hyperref[{fig:#2}]{%
    Fig.~\ref*{fig:#2}%
    \ifx\\#1\\%
    \else
      ~(#1)%
    \fi
  }%
}

\newcommand*{\Eqref}[2][]{%
  \hyperref[{eq:#2}]{%
    Eq.~\ref*{eq:#2}%
    \ifx\\#1\\%
    \else
      ~(#1)%
    \fi
  }%
}

\newcommand*{\Tabref}[2][]{%
  \hyperref[{tab:#2}]{%
    Table~\ref*{tab:#2}%
    \ifx\\#1\\%
    \else
      ~(#1)%
    \fi
  }%
}

\newcommand*{\Secref}[2][]{%
  \hyperref[{sec:#2}]{%
    Section~\ref*{sec:#2}%
    \ifx\\#1\\%
    \else
      ~(#1)%
    \fi
  }%
}

\begin{document}

\title{Data Processing Techniques for Ion and Electron Energy Distribution Functions}

\author{A. Caldarelli $^*$}
\affiliation{Te P\=unaha \=Atea - Auckland Space Institute, Department of Engineering Science, The University of Auckland, Auckland, 1010, New Zealand}
\author{F. Filleul $^*$}
 \email{felicien.filleul@auckland.ac.nz}
 \affiliation{Te P\=unaha \=Atea - Auckland Space Institute, Department of Physics, The University of Auckland, Auckland, 1010, New Zealand}
 \author{R.W. Boswell}
 \affiliation{Space Plasma, Power and Propulsion Laboratory, Research School of Physics, The Australian National University, Canberra, ACT 2601, Australia}
\author{C. Charles}
 \affiliation{Space Plasma, Power and Propulsion Laboratory, Research School of Physics, The Australian National University, Canberra, ACT 2601, Australia}
 \author{N.J. Rattenbury}
 \affiliation{Te P\=unaha \=Atea - Auckland Space Institute, Department of Physics, The University of Auckland, Auckland, 1010, New Zealand}
 \author{J.E. Cater}
 \affiliation{Te P\=unaha \=Atea - Auckland Space Institute, Department of Engineering Science, The University of Auckland, Auckland, 1010, New Zealand}
\date{\today}

\def\thefootnote{*}\footnotetext{These authors contributed equally to this work}

\begin{abstract}

Retarding field energy analyzers and Langmuir probes are routinely used to obtain ion and electron energy distribution functions (IEDF, EEDF). These typically require knowledge of the first and second derivatives of the $I$-$V$ characteristics, both of which can be obtained in various ways. This poses challenges inherent to differentiating noisy signals, a frequent problem with electric-probe plasma diagnostics. A brief review of commonly used analog and numerical filtering and differentiation techniques is presented, together with their application on experimental data collected in a radio-frequency plasma. The application of each method is detailed with regards to the obtained IEDF and EEDF, the deduced plasma parameters, dynamic range, energy resolution and signal distortion.

\end{abstract}

\maketitle

\section{Introduction}\label{sec:Intro}

Distribution functions offer a detailed description of plasma charged particles: from obtaining some useful plasma parameters (e.g. density, temperature and potential) to deducing heating and transport mechanisms, collision and reaction rates, as well as detecting the presence of particles sub-populations \cite{godyak1993probe,charles2004laboratory,takahashi2009transport,boswell2015non}. The ion-energy distribution function (IEDF) is extensively employed to characterize the ion kinetics in plasma processing techniques and in the development of spacecraft electric propulsion technologies, where the detection of accelerated ion populations, such as ion beams, is important \cite{charles2004laboratory,ingram1988ion,habl2020ion}. The electron-energy distribution function (EEDF) gives an extensive picture of the dynamics of electrons. This is essential for characterizing the plasma generation mechanisms in laboratory and commercial plasma devices, as well as in investigating fundamental aspects of plasma physics, e.g. the thermodynamics of magnetized electrons \cite{godyak1992evolution,takahashi2020thermodynamic}.

Electrostatic probes, such as retarding field energy analyzers (RFEA) and single Langmuir probes (LP), are some of the most common instruments for measuring ion and electron distribution functions, respectively. The plasma potential, and the ion beam potential and density are usually obtained through the first derivative of an RFEA $I$-$V$ characteristic. From a Langmuir probe $I$-$V$ characteristic, the floating potential, electron temperature, plasma density, and plasma potential can be inferred via the classical Langmuir method \cite{mott1926theory,merlino2007understanding}. While popular thanks to its apparent simplicity, this method is only valid for plasmas with Maxwellian electron distributions and can be inaccurate in determining the plasma potential and ion current \cite{godyak2011probe,godyak2021rf}. Alternatively, the Druyvesteyn method directly determines the electron distribution function from the second derivative of the LP $I$-$V$ characteristics, a process inherently more robust and information-rich than the Langmuir method, yet more challenging owing to the double differentiation \cite{druyvesteyn1930niedervoltbogen}.

The ion-energy distribution function has usually been computed by obtaining the first derivative of the collected current over the discriminator voltage \cite{Bohm1993, Charles2000, charles2004laboratory, Cox2008, takahashi2010analog, Bennet2018_RFEA, ingram1988ion,habl2020ion,Imai_2021}. It can be shown that, by assuming a one-dimensional velocity distribution \cite{Bohm1993}, the ion-energy distribution $g_{\rm i}(\varepsilon_{\rm i})$ is proportional to the negative derivative of the collected current $I_{\rm C}$ with respect to the discriminator grid voltage $V_{\rm D}$:
\begin{equation}
\label{eq:IEDF_eq}
	g_{\rm i}(\varepsilon_{\rm i}) = -\frac{1}{t^4 A_{\rm o} e^2} \sqrt{\frac{m_{\rm i}}{2 e V}} \frac{\diff I_{\rm C}}{\diff V_{\rm D}}~,
\end{equation}
where $t$ is the probe grid transmission factor (four grids in the case of the RFEA used in this work) \cite{charles2004laboratory}, $A_{\rm o}$ is the area of the probe orifice, and $m\textsubscript{i}$ is the ion mass. Thus, the IEDF can be calculated simply by differentiating the measured current with respect to the discriminator voltage. For a single Gaussian distribution, the voltage at which the IEDF peak occurs is defined as the local plasma potential, $V\textsubscript{p}$. If a population of accelerating ions is present, as in the case of a plasma expanding through a double-layer \cite{charles2004laboratory, Keese2005,Bennet2018_RFEA} and for the considered gas pressures (i.e., $\sim~$0.2 - 1$~$mTorr), the IEDF would show a second peak at a higher energy located at the ion beam potential, $V\textsubscript{B}$.

\citet{druyvesteyn1930niedervoltbogen} showed that for an isotropic plasma, the EEDF is proportional to the second derivative of the electron current from the $I$-$V$ characteristic of a Langmuir probe.
The relationship between the EEDF and the second derivative of the LP electron current $I_{\rm e}$ with respect to the biasing voltage $V_{\rm bias}$ is
\begin{equation}\label{eq:EEDFDIDV}
g_{\rm ed}(\varepsilon_{\rm e}) =  \frac{2\sqrt{2m_{\rm e} eV}}{A_{\rm p}e^3} \frac{\diff^2 I_{\rm e}}{\diff V_{\rm bias}^2}~.
\end{equation}
Here, $V = V_{\rm p} - V_{\rm bias}$ and $V_{\rm p}$ is the local plasma potential. $\varepsilon_{\rm e}$ is the electron energy in electronvolts, $A_{\rm p}$ is the probe area, $e$ is the elementary charge and $m_{\rm e}$ the electron mass. The electron probability distribution function (EEPF) is then $g_{\rm ep}(\varepsilon_{\rm e}) = g_{\rm ed}(\varepsilon_{\rm e}) / \sqrt{\varepsilon_{\rm e}} $. In an isotropic electron gas, the EEDF and EEPF contain the same information as the electron distribution function (which includes the velocity space). The local plasma potential $V_{\rm p}$ is found from the zero crossing of $\diff^2I_{\rm e}/\diff V_{\rm bias}^2$. The the electron density can be retrieved from
\begin{equation}\label{eq:ne}
n_{\rm e} = \int_{0}^{\infty} g_{\rm ed}(\varepsilon_{\rm e}) \diff \varepsilon_{\rm e} ~,
\end{equation}
and the effective electron temperature $T_{\rm eff}$ can be calculated from the average electron energy $<\varepsilon_{\rm e}>$ as
\begin{equation}\label{eq:Teff}
T_{\rm eff} = \frac{2}{3}<\varepsilon_{\rm e}> = \frac{2}{3} \int_{0}^{\infty} \varepsilon_{\rm e} g_{\rm ed}(\varepsilon_{\rm e}) \diff \varepsilon_{\rm e} ~.
\end{equation} 
Plotting the natural logarithm of the EEPF can show a departure from a Maxwellian energy distribution function if the slope of the EEPF is not linear, e.g. bi-Maxwellian or Druyvesteynian. If the EEPF is Maxwellian, the electron temperature can also be deduced from the slope of the natural logarithm of the EEPF and is equal to $T_{\rm eff}$. 

IEDF and EEDF are usually obtained through a variety of analog and numerical differentiation routines. Since measurement noise amplification is inherent to the differentiation process, filtering/fitting methods may be necessary to smooth the measured $I$-$V$ data without causing significant distortion and to obtain accurate first and second derivatives. Among the numerical methods, the Savitzky-Golay filter is one of the most commonly employed for IEDF/EEDF derivations \cite{Savitzky1964,sudit1993workstation,bechu2013multi,Conde2017,Damba2018}. \citet{fernandez1995new} implemented a Gaussian filter to smooth the $I$-$V$ characteristics and evaluate the electron-energy distribution function. This method was compared with the Savitzky-Golay filter and the B-spline approximation methods. \citet{magnus2008digital} compared the use of the Savitzky-Golay filter, as well as the Gaussian filter, the Blackman window and polynomial fitting, to smooth both simulated and experimental $I$-$V$ curves. The mean squared error, the correlation coefficient and the residuals of the different methods were compared, coupled with a visual assessment of the EEPF. Alternatively, analog differentiation has been performed by using appropriate electronics circuitry to obtain the IEDF and the EEPF \cite{sloane1934xv,godyak1992measurement,takahashi2009transport,takahashi2010analog,Yadav2018}. The IEDF has also been obtained using a Gaussian deconvolution method \cite{charles2004laboratory, Cox2008, Imai_2021}.
 
The aim of this paper is to review different signal processing and differentiation methods for the acquisition of optimal ion- and electron-energy distribution functions. The study discusses analog and numerical processing techniques that are routinely used to obtain EEDFs and IEDFs. With respect to the latter, the focus is on processing IEDFs that show the presence of multiple ion populations. The different numerical processes reviewed to obtain the first and second derivatives from the I-V characteristics are: Savitzky-Golay (SG) filter, B-spline piecewise polynomial fitting (BS), Gaussian filter (GF), and the Blackman window filter (BW). The ion-energy distribution function is also computed using the Gaussian deconvolution method. The analysis presented is intended as a baseline approach for the manipulation of $I$-$V$ characteristics obtained from experimental data in a plasma discharge. Specifically, data collected in a magnetized radio-frequency plasma discharge at low gas pressures are used, which contain a higher noise level than, e.g. DC plasma discharges. This work gathers the lessons learned in the acquisition and processing of electric probe data to ease the reader's process of obtaining charged-particles energy distribution functions of suitable quality.

The work is organized as follows; \Secref{Appara} describes the experimental set-up, including the plasma diagnostic probes. \Secref{DataProc} introduces the analog and numerical filtering/differentiation methods employed. Finally, \Secref{ResDes} provides an example application of data processing analysis for ion- and electron-energy distribution functions obtained from the different methods starting from raw experimental data.

\section{Apparatus \& Diagnostics}\label{sec:Appara}

\subsection{Apparatus}

The data used in this study has been acquired using \textit{Moa} \cite{Caldarelli2022JEP}, a radio-frequency plasma reactor that is an expansion of \textit{Huia} \cite{filleul2021characterization,filleul2022ion}. It consists of a $150~$cm long, $9~$cm inner diameter borosilicate glass tube plasma source connected to a $70~$cm long, $50~$cm diameter steel expansion chamber hosting the pumping system and the pressure gauges. The location of the interface between the plasma source and the expansion chamber is defined as $(r,z) = (0,0)~$cm. A base pressure of $\sim 10^{-7}~$Torr is routinely achieved, while the argon working pressure ranges between $0.5\times 10^{-3}~$Torr and $1 \times 10^{-3}~$Torr. The plasma is triggered and energized in the plasma source using a $1$-$1/3$ turns loop antenna connected through an L-type matching network to a $1~$kW RF generator working at $27.12~$MHz. A static magnetic field is applied with a pair of movable Helmholtz solenoids, placed concentrically with the glass tube. The coils position along the plasma source length can be adjusted to obtain the desired magnetic field topology.

\subsection{Retarding Field Energy Analyzer}

The retarding field energy analyzer used to measure the ion-energy distribution function consists of four mesh grids (namely, earth, repeller, discriminator, and secondary grid) and a nickel collector plate that collects the incoming ion flux. The grids are made of a nickel mesh with a transmission factor of 59\% which are attached to a copper support. The probe orifice has a diameter of $2~$mm, and a $0.1~$mm thick polyimide sheet is used to electrically insulate the grids and the collector plate. The design of this probe has been extensively used in similar experiments \cite{Charles2000,charles2004laboratory,Cox2010,Bennet2018_RFEA}, and a detailed description of the RFEA and its electric circuit used herein are presented in Ref.\,\onlinecite{caldarelli2021preliminary}. In order to map the energy of the ions, the voltage of the discriminator grid is swept from $0~$V to $80~$V to obtain the $I$-$V$ characteristics. It has to be noted that from RFEA measurements, only the energy distribution function of the ions that fall through the probe sheath within an acceptance angle of $\pm 40^\circ$ are obtained \cite{charles2004laboratory}. When taking measurements of the ion current in an RF plasma with an RFEA, the collected data can be affected by different mechanisms. In particular, in an RF excited plasma, broadening of the energy distribution caused by RF modulation can be observed \cite{Conway_1998, Charles2000, charles2004laboratory} that could lead to peak separation, giving false data on the presence of different ion populations.
Additionally, possible ion-neutral collisions (elastic and charge-exchange) occurring in the probe sheath and inside the analyzer could also affect the collected ion energy. However, for the low-pressure case analyzed, collisions inside the RFEA and in the sheath can be considered negligible, i.e. $\lambda_{\rm i} >> s, d$ (where $s$ is the sheath width in front of the RFEA and $d$ is the repeller-discriminator grid distance).

\subsection{RF Compensated Langmuir Probe}

As from Langmuir's work, sizing a single Langmuir probe requires the plasma volume disturbed by the probe to be much smaller than the electron mean free path $\lambda_{\rm e}$ \cite{godyak2011probe}. This ensures that the probe only disturbs the plasma in ways that can be accounted for by theory. Following recommendations in Ref.\,\onlinecite{godyak2021rf}, Langmuir's theory is satisfied if the probe tip radius $r_{\rm p}$, the tip length $l_{\rm p}$, the tip holder radius $r_{\rm h}$ and the local Debye length $\lambda_{\rm D}$ all satisfy the condition $r_{\rm p} \ln{\left( \pi l_{\rm p} / 4 r_{\rm p} \right)}, r_{\rm h}, \lambda_{\rm D} \ll \lambda_{\rm e}$. Moreover, to guarantee Druyvesteyn's isotropic plasma condition in the presence of an applied magnetic field, the probe tip should be kept perpendicular to the streamlines and the tip radius be much smaller than the electron Larmor radius \cite{godyak2011probe}. To meet these conditions, the LP used in this work is made from a tungsten tip with $r_{\rm p}=25~\mu$m and $l_{\rm p}=3~$mm, mounted at the extremity of a glass pipette tapering down to $r_{\rm h}=0.75~$mm. 

Measuring an EEDF requires biasing the LP above the local plasma potential and can typically draw electron currents in the order of a few milliamperes. Therefore, in order to close the LP electric path and to ensure that all the biasing voltage is developed across the probe sheath, the existence of a low impedance sheath at conductive walls should be verified (see Ref.\,\onlinecite{godyak2011probe} for more details). For this purpose a grounded aluminium sleeve of adequate surface area was inserted inside the dielectric plasma source tube at the end opposite to the diffusion chamber \cite{filleul2021characterization}.

\begin{figure}[!ht]
\centering
\includegraphics[width=6.cm]{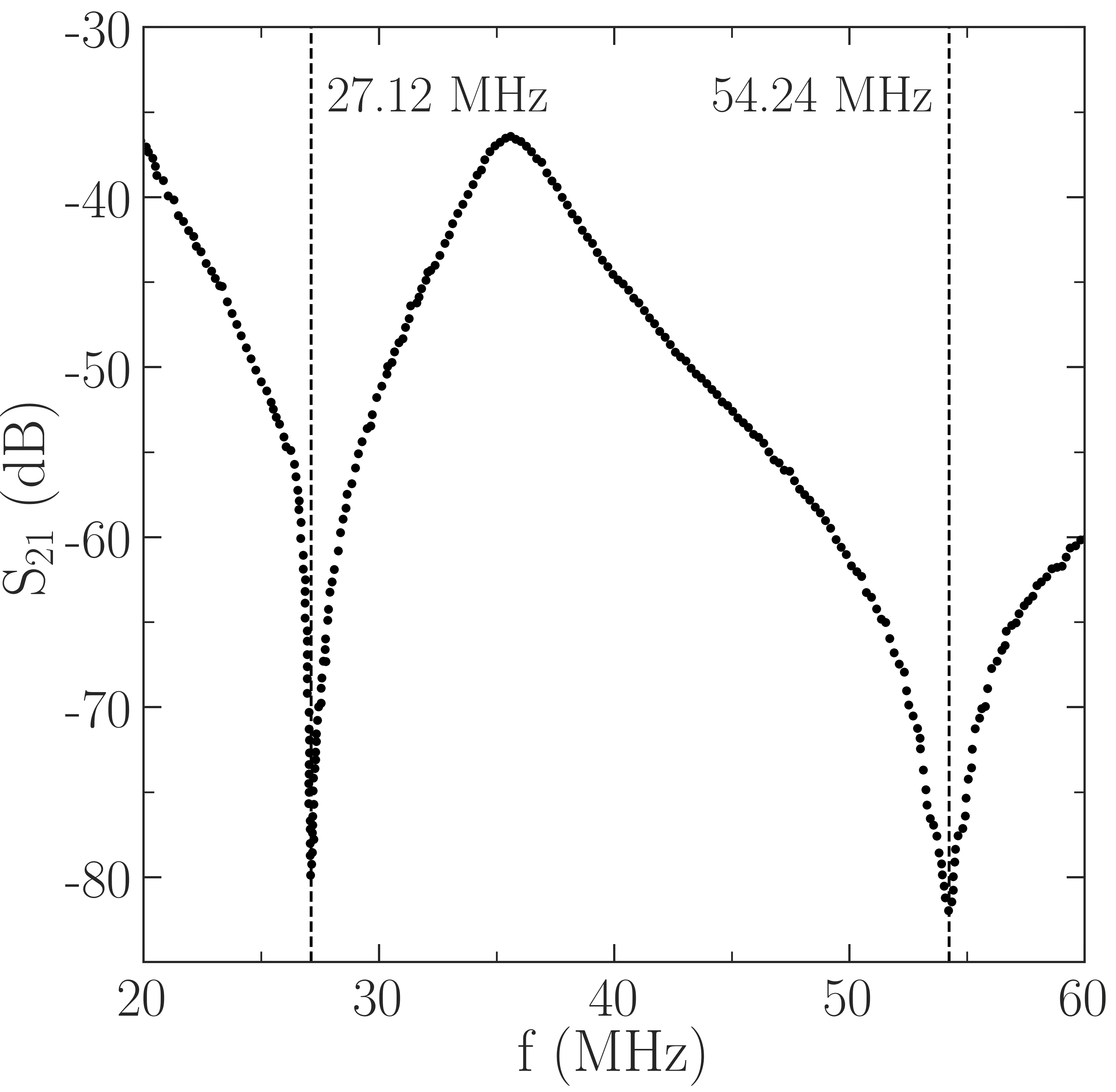}
\caption{The attenuation spectrum of the RF compensated probe used in this study.}
\label{fig:CPSpectrum}
\end{figure}

With the loop antenna capacitive coupling to the plasma, oscillations of the plasma potential at the applied radio-frequency and its harmonics are known to cause severe distortions to LP $I$-$V$ characteristics \cite{sudit1994rf,godyak2011probe}. While this effect is not easily visible in the probe $I$-$V$ characteristic, it becomes evident with the second derivative and, without proper mitigation, it severly compromises EEDF studies. Because the root-mean-square of $V_{\rm p}$ at the first and second harmonics are larger than $T_{\rm e}/2$ in the present apparatus, RF compensation was added to the Langmuir probe, following design steps described in detail in \citet{sudit1994rf} and \citet{godyak2011probe}. Four resonant chokes (\textit{TE Connectivity SC30100KT}) and a reference electrode were incorporated in the probe head. The chokes self-resonant frequencies (SFR) were tuned to $27.12~$MHz and $54.24~$MHz with shunt capacitors of approriate values. The filter is housed inside the glass pipette, placed as close as possible to the probe tip to minimize stray capacitance. Shown in \figref{CPSpectrum}, the attenuation spectrum of the completed probe is checked with a spectrum analyzer for eventual detuning induced by the housing and adjacent wiring. The achieved attenuation of at least $-80~$dB at the first and second harmonics provides sufficient filter impedance to suppress the $V_{\rm p}$ RF oscillations. The reference electrode, connected to the probe tip through a capacitor, reduces the probe sheath impedance to ensure that the measuring tip follows the the plasma potential RF oscillations. For chokes containing ferrite cores, care should be taken to ensure that the ferrite does not saturate under the applied magnetic field or else the SFR will be shifted. The chokes employed were confirmed to be unaffected at the applied magnetic field strength of 300$~$G. The maximum probe biasing voltage was adjusted to $V_{\rm bias} \simeq V_{\rm p} + 5~$V to avoid excessive electron current collection which could damage the probe tip or modify its work function. Tip contamination effects were monitored by checking for hysteresis in the up and down sweeps of $V_{\rm bias}$. Probe cleaning by ion bombardment was used when necessary.

\subsection{Data Collection}

The Langmuir probe is mounted on a movable doglegged eccentric dielectric shaft which can span the entire length of the plasma source tube \cite{filleul2021characterization}. The RFEA is mounted on a movable (rotation and translation) 6.35$~$mm diameter steel shaft introduced in the top port of the expansion chamber, with the probe orifice facing the plasma source exit to detect any possible directional ion beam. The LP current is deduced by measuring the voltage drop across a resistor with an isolation amplifier. The output of the amplifier is then split and fed into both a data acquisition system (DAQ) for digitization, and into an analog differentiator which outputs the first and second time derivatives. The LP $V_{\rm bias}$ and the RFEA $V_{\rm D}$ are defined by several periods of a triangle wave function. The data acquisition is triggered by the output on the function generator to ensure signal synchronicity and the voltages are digitized at $4,000$ samples per period. One LP measurement is the average of $100$ $V_{\rm bias}$ sweeps, while one RFEA measurement averages $200$ $V_{\rm D}$ sweeps.

\section{Data Processing Methods}\label{sec:DataProc}
Obtaining an accurate IEDF or EEDF requires choosing between different techniques to compute the first and second derivative of the $I$-$V$ characteristics. An ideal method would provide an undistorted derivative with a signal-to-noise ratio (S/N) equal or lower to the raw data S/N. However, in practice there is a trade-off between noise level reduction and data distortion. Fortunately, analog and numerical methods can typically combine both noise filtering and differentiation of the raw $I$-$V$ data. Some of the most commonly used numerical routines are presented in this section: the Savitzky-Golay filter, the B-spline fitting, the Gaussian and the Blackman filters, together with an analog differentiation technique. The Gaussian deconvolution method to obtain IEDFs is also described.

\subsection{Analog Differentiation}

The use of analog differentiator circuits predates numerical differentiation methods, and they are still regularly employed for obtaining EEDFs and occasionally for IEDFs \cite{godyak1992measurement,takahashi2009transport,takahashi2010analog,boswell2015non,Yadav2018}. The active analog differentiator used in this study, designed to work with a sweeping frequency of $10~$Hz, follows the design described in \citet{takahashi2010characterization}. Two cascading operational amplifiers (\textit{Renesas CA3140}) are each tuned to resonate at ${\small \sim} 800~$Hz by adjusting the resistors and capacitors values. This results in a linear gain frequency increase of $20~$dB per decade up to the resonance frequency, i.e. input signals of frequencies up to ${\small \sim}800~$Hz are time-differentiated. Beyond the resonance frequency, the circuit acts as an integrator as the gain linearly decreases until becoming less than unity at ${\small \sim} 80~$kHz. Therefore the analog differentiator configuration also works as a band-pass filter, attenuating high frequency noise. Since real-world circuits deviate from the ideal-case, the linearity was checked to be strictly valid up to at least ${\small \sim}150~$Hz. Each stage of the analog differentiator gives an output proportional to the first and second time derivative of the probe $I(t)$ signal. Knowing the biasing/discriminator voltage sweep period and voltage swing, the differentiator output can be multiplied by $\diff t / \diff V$ to retrieve the $\diff I_{\rm p} / \diff V$ derivative. Differentiator circuits have the advantage of producing repeatable outputs since no tuning of parameters for different experimental conditions are needed. This simplifies and speeds up the data processing.

\subsection{Polynomial Methods}

A polynomial least-squares regression to model the data using a single polynomial of degree $N$ via a least squares method has been previously attempted, giving poor results \cite{magnus2008digital}. Since the $I$-$V$ characteristics of interest can be visualized as part exponential and part linear, a single polynomial regression would require a high order to fit the data to a satisfactory level, e.g. $N > 20$, with $N$ the polynomial degree \cite{magnus2008digital}. This leads to Runge's phenomenon and over-fitting \cite{Celant2016Interpolation}, which is visible in the results presented by \citet{magnus2008digital}. Another shortcoming is the non-locality of the polynomial fit, i.e. distant data points directly impact the local fit \cite{perperoglou2019review}. The SG filter and the B-spline methods both circumvent these problems by fitting low degree polynomials to subsets of the data.

\subsubsection{Savitzky-Golay Filter}
The Savitzky-Golay filter is frequently used for smoothing and differentiating current-voltage characteristics of Langmuir probes and RFEAs \cite{Xu2012_SGapplication,bechu2013multi,Gulbrandsen2017,Damba2018}. The filtering is achieved by doing a running fitting of a polynomial of degree $N$ using the least-squares method on a subset of the $(x,y)$ experimental data of width $M > N$ centered around a given point $x_i$ (forcing $M$ to strictly be an odd number). Evaluating the resulting polynomial at $x_i$ gives the smoothed value $Y_i$ and the process is repeated for $x_{i+1}$ etc. Owing to the nature of this process, the first and last $\frac{M-1}{2}$ data points should be disregarded or the original data-set extrapolated by $\frac{M-1}{2}$ extra points at both ends. 

\citet{Savitzky1964} showed that for equally spaced data points, an analytical solution to the least-squares polynomial smoothing exists. Each one of the $N+1$ coefficients of the least-squares polynomial fitting is computed as a linear combination of the $y$ data points inside the subset of width $M$. It was further demonstrated that the coefficients of the linear combinations are function of $(N,M)$ only. These can be tabulated for all $(N,M)$ pairs and applying the filter only requires convolution of the coefficients with the raw data \cite{Savitzky1964}. A benefit of this property is that the smoothed signal can be differentiated $N$ times by convolution with the $N^{th}$ derivative of the fitting polynomial, i.e. with the appropriate convolution coefficients \cite{luo2005properties}. This avoids having to recourse to numerical finite difference methods which tend to decrease the S/N. This can be understood by considering that the differential operator in the frequency domain is proportional to the frequency itself. For a given $M$, the pairs $N=0-1$, $2-3$, $4-5$, etc. give the same convolution coefficients for smoothing the raw data and evaluating the even derivatives, while $N=1-2$, $3-4$, $5-6$ give the same coefficients for odd derivatives \cite{luo2005properties}. Therefore for a given $M$, using $N=2$ gives an identically smoothed $2^{\rm nd}$ derivative as using $N=3$.

The SG filter acts as a low-pass filter with a flat passband whose cut-off frequency is a function of $N$ and $M$. The cut-off frequency typically increases with $N$ and decreases with $M$ \cite{luo2005properties}. For large $M$, the benefit of lower cut-off frequencies is compromised by distortions of the higher frequency content of the data. On the other hand, a large $N$ can preserve the narrow features of a signal (such as peaks) but at the cost of increasing the cut-off frequency \cite{Press1990}. Obtaining an optimal filtering therefore requires an iterative process and a trade-off between the values of $N$ and $M$. 

\subsubsection{B-spline Fitting}
The piecewise polynomial approach used for data smoothing is the cardinal or uniform B-spline method \cite{de2001practical}. This polynomial fitting method avoids Runge's phenonenon by making use of a piecewise approach: the data is split in $k+1$ pieces joined by $k$ equidistant points called knots. Like other polynomial regressions, the fitting polynomial $f(x)$, or spline function, of degree $N$ can be expressed as a linear combination of coefficients $\beta_{i}$ and $k + N + 1$ basis functions (polynomials) $B_{i,N}(x)$:
\begin{equation}
	f(x) = \sum_{i} \beta_i B_{i,N}(x)~.
\end{equation}
Spline functions can therefore be calculated on different bases and the B-spline is a particular case for basis functions with local support, i.e. taking non-zero values only inside the interval on which they are defined, which spans $N+2$ knots. As a result, this method results in a higher numerical stability compared to other splines \cite{perperoglou2019review}. The spline function $f$ is fitted to the data by using the least-square principle to calculate the $k + N + 1$ coefficients $\beta_{i}$. 
Additionally, the continuity conditions need to be verified, such that the function is $N-1$ times differentiable at the knots and that $f''= f'''= 0$ at the endpoints. 
The larger the number of knots, the higher the accuracy of the data fitting. However, more knots may result in over-fitting and/or poor noise rejection. A small number of knots could also over-smooth the data. As for the SG filter, the first and second derivatives of $f$ can be analytically derived.

\subsection{Window Filters}

\subsubsection{Gaussian Filter}
The Gaussian filter was developed by \citet{Hayden1987} who introduced a new smoothing routine involving the use of a Gaussian distribution function as a filter. \citet{fernandez1995new} and \citet{magnus2008digital} used this algorithm to smooth the $I$-$V$ characteristics measured with a Langmuir probe and obtain the electron-energy probability function. The working principle of the Gaussian filter is based on the assumption that the response function of the instrument and its electronics can be approximated as a Gaussian distribution with a given standard deviation $\sigma$:
\begin{equation}
	g(x) = \frac{1}{\sigma \sqrt{2 \pi}} \exp{\left( -\frac{x^2}{2 \sigma^2} \right)}~.
\end{equation}
The measured signal can then be calculated from the convolution of the experimental data $d$ with $g$ ($d \ast g$). Because the noise is not correlated with the instrument response function, $g$ can be used for noise suppression. Owing to the property $ (d \ast g)' = d' \ast g = d \ast g'$, the raw data can be convolved with the first and second derivative of the Gaussian function in order to obtain the first and second derivative of the $I$-$V$ characteristic instead of using the central difference method. Nevertheless, for the sake of comparison with earlier results \cite{magnus2008digital}, the central difference method was used to evaluate the derivatives.

\subsubsection{Blackman Window Filter}
The Blackman window $w$ is a standard moving average window, having zero value outside a chosen interval. It has been previously used to filter $I$-$V$ characteristics in EEPF applications by \citet{magnus2008digital} and \citet{ROH20151173}. The Blackman window is defined by only one filtering parameter, i.e. the window size $M$:
\begin{equation}
\begin{split}
    w(n) = 0.42 - 0.5 & \cos\left(\frac{2\pi n}{M}\right) + 0.08 \cos\left(\frac{4\pi n}{M}\right), \\
    & 0 \leq n \leq M ~.
\end{split}
\end{equation}
It is characterised by the highest reduction in the sidelobe level (down to $\approx -58~$dB) when compared to the other window functions (e.g. Hanning) \cite{ChassaingRulph2005Dspa}. However, it has a wider mainlobe resulting in a less sharp transition band in the passband of the filter that can be improved by choosing a wider window size. The derivatives cannot be obtained analytically for the Blackman window, thus a central difference operator was used after filtering the raw $I$-$V$ data.

\subsection{Gaussian Deconvolution}
The Gaussian deconvolution method, or Gaussian fitting, has been used to obtain the ion-energy distribution function from the $I$-$V$ characteristics measured by a retarding field energy analyzer \cite{charles2004laboratory, Cox2008, Imai_2021}. This involves an iterative process of integrating a Gaussian curve that models the ion-energy distribution function until the raw $I$-$V$ data are reconstructed. The fitting can either be manual, or an automatic Gaussian fitting algorithm can be implemented that minimises the sum of the squared errors between the measured and the reconstructed data.

For a single ion population, the ion-energy distribution function can be modeled with a Gaussian curve
\begin{equation}
    g(V_{\rm D}) = a e^{-\left(\frac{V_{\rm D} - V_{\rm peak}}{c}\right)^2},
\end{equation}
where the parameters to be fit are: the amplitude $a$, the curve width $c$, and the location of the Gaussian peak $V_{\rm peak}$. When a second ion population is present, as in the case of an accelerated ion beam in a collisional plasma, the IEDF requires the fitting of two or three Gaussian curves which are summed to accurately model the measured $I$-$V$ curve \cite{charles2004laboratory, Bennet2018_RFEA}. An advantage of this technique is that it does not involve any filtering or differentiation steps since the IEDF is obtained in a retrograde approach.

\section{Example of Application on a Data Set}\label{sec:ResDes}

\subsection{Optimization of Numerical Methods}\label{sec:Stats}

The optimal evaluation of the ion and/or energy distribution functions through numerical methods requires a compromise in the choice of parameters to achieve sufficient noise-reduction without causing significant distortion of the fitted $I$-$V$ characteristics that could lead to large errors in the inferred plasma parameters and loss of information. 

An analysis similar to that used by \citet{fernandez1995new} and \citet{magnus2008digital} is used to assess the numerical methods in this work. It is noted that in Ref.\,\onlinecite{fernandez1995new} and Ref.\,\onlinecite{magnus2008digital}, the comparison of the different techniques was conducted on simulated $I$-$V$ characteristics, while in this study only experimental data are used. The following parameters, evaluated between the raw probe collected current $I_{\rm p}$ and the processed data $I_{\rm f}$, were considered: the mean squared error (MSE) and the Shapiro-Wilk (SW) test statistic $W$. 

When comparing different regression models, an optimal processing technique would return the lowest mean squared error. However, because the MSE value can range between zero and any larger number depending on the scale of the variables, it does not provide a standardized metric to assess how well the processing actually models the observed data. Moreover, the MSE on its own does not provide a way to highlight under-smoothing and over-fitting as these situations could return the lowest possible MSE. Thus, an analysis of the distribution of the residuals ($I_{\rm p} - I_{\rm f}$) is necessary to complement the MSE. In particular, verifying that the residuals are normally distributed ensures that the data processing does not significantly distort the $I$-$V$ features \cite{SenAshishK1990_regression}. The Shapiro-Wilk test checks the null-hypothesis that the residuals are normally distributed by comparing the computed statistic $0 < W < 1$ with tabulated values \cite{shapiro1965analysis,rahman1997modification}. For a given significance level, if $W$ is smaller than the tabulated value then the null-hypothesis can be rejected, i.e. the residuals are not normally distributed. Since filtering and fitting always somewhat distorts the data, one should aim to maximize $W$ and re-evaluate the data processing when the null-hypothesis is rejected. Taking a 0.01 significance level and given the samples size used in this study, the null-hypothesis is rejected if $W<0.996$ \cite{rahman1997modification}. Therefore, a parametric analysis is conducted together with a visual optimization of the IEDF and EEPF for determining the optimal parameters for each numerical method. The assessment requires a trade-off between minimizing the MSE, maximizing $W$ and minimizing alterations of the raw $I$-$V$ characteristic features.

\begin{table}[!ht]
\renewcommand{\arraystretch}{1.0}
\caption{Mean-squared error and Shapiro-Wilk test for different values of degree $N$ and number of knots $k$ of the B-spline applied to the RFEA data in \figref{IEDF_parameters_BS}.}
\label{tab:IEDF_parameters_BS}
\centering
\setlength{\tabcolsep}{0.5em} 	
\begin{tabular}{c c c c} \toprule	
       $N$ & $k$ & MSE$ / 10^{-4}~\rm(\mu A^{2})$ & $W$\\
       \midrule
        3 & 5 & $ 45.62$ & 0.9935\\
        3 & 15 & $ 9.25$ & 0.9989\\
        3 & 35 & $ 8.92$ & 0.9985\\
        3 & 80 & $ 8.87$ & 0.9985\\
       \midrule
        2 & 15 & $ 9.42$ & 0.9987\\
        4 & 15 & $ 9.75$ & 0.9986\\
        \bottomrule
\end{tabular}
\end{table}

\begin{figure}[!ht]
    \centering 
    \includegraphics[width=8.5cm]{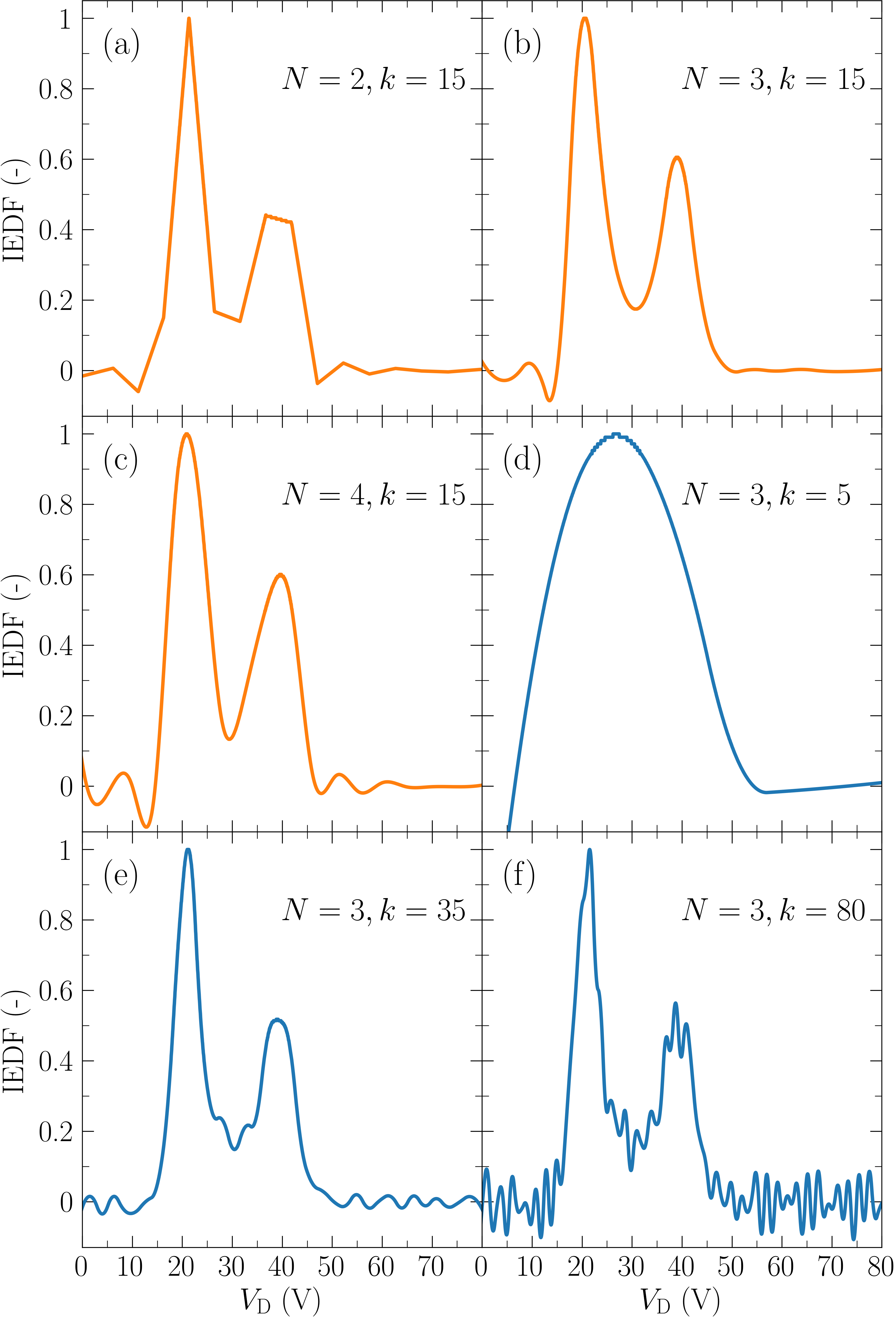}
    \caption{Normalized first derivative of the I-V characteristics measured by the RFEA obtained
with the B-Spline polynomial for various values of degree $N$ (a)-(c), and number of knots $k$ (d)-(f).}
    \label{fig:IEDF_parameters_BS}
\end{figure}

Specifically, the optimization of the IEDFs involves the careful computation of the first and second peak of the energy distributions, while achieving enough noise reduction when evaluating $\diff I_{\rm C}/\diff V_{\rm D}$. An excessive smoothing of the $I$-$V$ characteristics would modify the voltage at which the $I$-$V$ curves present the gradient changes that corresponds to the center of two ion populations present. In turn, this would yield an incorrect evaluation of the plasma and ion beam potentials, and an inaccurate derivation of the ion-energy distribution.
The dependency of the choice of the numerical processing parameters on the quality of the ion-energy distribution function was studied for each numerical method. \figureref{IEDF_parameters_BS} shows an example of the effect of the parameter choices on the first derivative for different values of polynomial degree $N$ and number of knots $k$ using the B-spline technique. The mean-squared error and the Shapiro-Wilk test are evaluated for each case, and the values are reported in \Tabref{IEDF_parameters_BS}. When analyzing the dependence of the fitted polynomial degree $N$ while $k=15$ is kept constant, the MSE is seen to increase with $N$, $W$ to decrease and the floor ripple level to increase (the Runge phenomenon). Since the $I_{\rm f}$ obtained from a B-spline of degree $N$ is $N-1$ continuously differentiable, a quadratic B-spline does not provide a smooth first derivative (see \figref[a]{IEDF_parameters_BS}). By extension, $N=4$ is the lowest B-spline degree that can be used to obtain a smooth second derivative for the EEPF.
If the number of knots is too small the evaluated IEDF is over-smoothed and the features of the $I$-$V$ characteristic are distorted ($W < 0.996$ for $N=3$ and $k=5$). In contrast, as shown in \figref[f]{IEDF_parameters_BS}, too high a number of knots causes an insufficient smoothing of the noise, which could make it difficult to determine the location of the peaks. For the data analyzed in this study, a degree of $N=3$ and a number of knots $k = 15$ are chosen for the B-spline polynomial since they result in the optimal IEDF i.e., lowest curve distortion and sufficient noise reduction.

\begin{figure}[!h]
\centering 
\includegraphics[width=8.4cm]{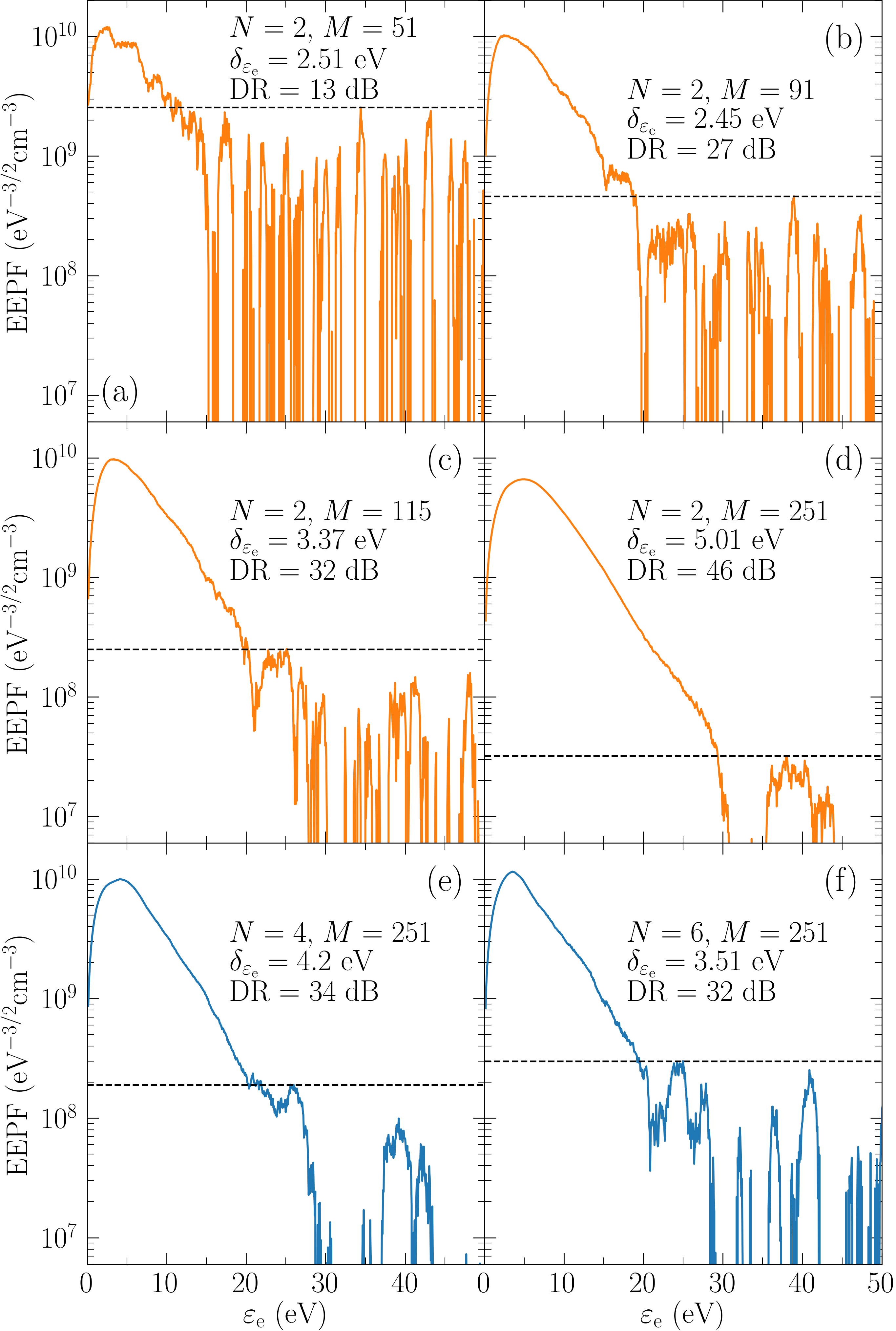}
\caption{EEPFs obtained from applying the Savitzky-Golay filter on an LP $I$-$V$ curve for $N=2$ and increasing window size $M$ (a)-(d) and $M=251$ with increasing value of $N$ (e)-(f). The horizontal dashed lines mark the minimum resolvable values of the EEPFs.}
\label{fig:SG_NumParamVar}
\end{figure}

\begin{table}[!h]
\renewcommand{\arraystretch}{1.0}
\caption{Mean-squared error (MSE) and Shapiro-Wilk test ($W$) for various values of degree $N$ and window size $M$ of the Savitzky-Golay applied to the LP data shown in \figref{SG_NumParamVar}.}
\label{tab:SG_NumParamVar}
\centering
\setlength{\tabcolsep}{0.5em} 	
\begin{tabular}{c c c c} \toprule	
       $N$ & $M$ & MSE$ / 10^{-4}~\rm(mA^{2})$ & $W$\\
       \midrule
        2 & 51 & $ 8.17$ & 0.9973\\
        2 & 91 & $ 8.50$ & 0.9961\\
        2 & 115 & $ 9.08$ & 0.9964\\
        2 & 251 & $ 4.37$ & 0.7135\\
        \midrule
        4 & 251 & $ 1.16$ & 0.9132\\
        6 & 251 & $ 8.86$ & 0.9937\\
     \bottomrule
\end{tabular}
\end{table}

Similarly, optimal evaluation of the EEDFs requires simultaneously maximizing the energy resolution and the dynamic range (DR). The energy resolution is characterized by the low-energy depletion $\delta_{\varepsilon_{\rm e}}$ which is defined as the $\varepsilon_{\rm e}$ value of the maximum of the EEDF. Because the bulk of the electrons are in the low-energy region of the EEPF, the larger $\delta_{\varepsilon_{\rm e}}$, the larger the errors in determining $n_{\rm e}$ and $T_{\rm eff}$ from \Eqref{ne} and \Eqref{Teff}. An EEDF is considered to have sufficient energy resolution if $\delta_{\varepsilon_{\rm e}} \leq (0.3-0.5)T_{\rm e}$ \cite{godyak2011probe,godyak2021rf}. The dynamic range of the EEPF is defined as the ratio of its maximum to its minimum resolvable value. A dynamic range $\geq 60~$dB (equivalently of 3 orders of magnitude and above) is necessary to resolve high-energy electrons in the inelastic range of the EEPF, which are detrimental in investigating excitation, ionization, and transport processes \cite{godyak2011probe,godyak2021rf}. An EEDF is often separated into its elastic and inelastic ranges, defined as $\varepsilon_{\rm e} < \varepsilon_{\rm e}^*$ and $\varepsilon_{\rm e} > \varepsilon_{\rm e}^*$, respectively. $\varepsilon_{\rm e}^*$ is the lowest excitation energy of the working gas, e.g. 11.55$~$eV for argon \cite{puech1986collision}. Minimizing $\delta_{\varepsilon_{\rm e}}$ and maximizing the DR is therefore paramount in the process of determining the optimal numerical parameters, together with minimizing the MSE and maximizing the value of $W$. 

\figureref{SG_NumParamVar} shows an example of the impact of varying the polynomial degree $N$ and moving window size $M$ of the Savitzky-Golay filter when applied to a compensated LP $I$-$V$ characteristic acquired with the experimental apparatus. For each $(N,M)$ pair, the corresponding MSE and Shapiro-Wilk test statistic $W$ are given in \Tabref{SG_NumParamVar}. In \figref[a)-(d]{SG_NumParamVar} $N=2$, and the effect of increasing $M$ is shown. As expected for an SG filter, increasing $M$ is beneficial from the dynamic range point-of-view since it improves from 13$~$dB to 46$~$dB owing to the decreasing cut-off frequency. However, this is accompanied by distortions of the signal, visible from the worsening of the low-energy resolution $\delta_{\varepsilon_{\rm e}}$, the MSE and $W$. Fixing $M$ and increasing $N$ to the next even integers (since $N=2$ and $N=3$ produce the same output), the signal distortion improves again at the cost of decreasing the DR, as shown in \figref[d)-(f]{SG_NumParamVar} and \Tabref{SG_NumParamVar}. Regarding the Shapiro-Wilk test, all the cases with $M=251$ failed the normality test with $W < 0.996$. Overall, trading-off between $\delta_{\varepsilon_{\rm e}}$, the DR, the MSE and $W$, the best performing filter for this data is for $(N=2,M=115)$. It is noted that none of the $(N,M)$ pairs produced an EEPF fulfilling the aforementioned requirements, since the dynamic range never reaches 3 orders of magnitude. Different $(N,M)$ pairs can be used whether the goal of obtaining the EEPF lies in the study of elastic or inelastic processes. Alternatively, the outputs of two SG filter applications could be combined into a high resolution and high dynamic range numerically derived EEPF, with the drawback of a more complicated post-processing \cite{ROH20151173}.

\subsection{Ion-Energy Distribution Function (IEDF)}

As mentioned in \Secref{Intro}, the ion-energy distribution function can be derived from the first derivative of the measurements acquired with a retarding field energy analyzer. The respective $I$-$V$ characteristics was taken in the expansion chamber with the RFEA orifice located 19$~$cm downstream of the magnetic nozzle ($z_{\rm P} = 19$\,cm). The experimental conditions were maintained at an RF power of $250~$W, argon pressure in the chamber of $0.5~$mTorr, and a maximum magnetic field on axis of $B_{z\rm ,max} = 330~$G. Under these conditions, the plasma density peaks axially under the solenoids reaching a value of $\approx ~ 10^{12}~\rm cm^{-3}$ which then decreases to $\sim ~ 10^{10}~\rm cm^{-3}$ in the expansion chamber.
The data processing of the collected current $I_{\rm C}(V_{\rm D})$ measured with the energy analyzer and the subsequent computation of the ion-energy distribution function followed the approach described in \Secref{Stats}.

\begin{figure*}[t]
\centering
\includegraphics[width=14.cm]{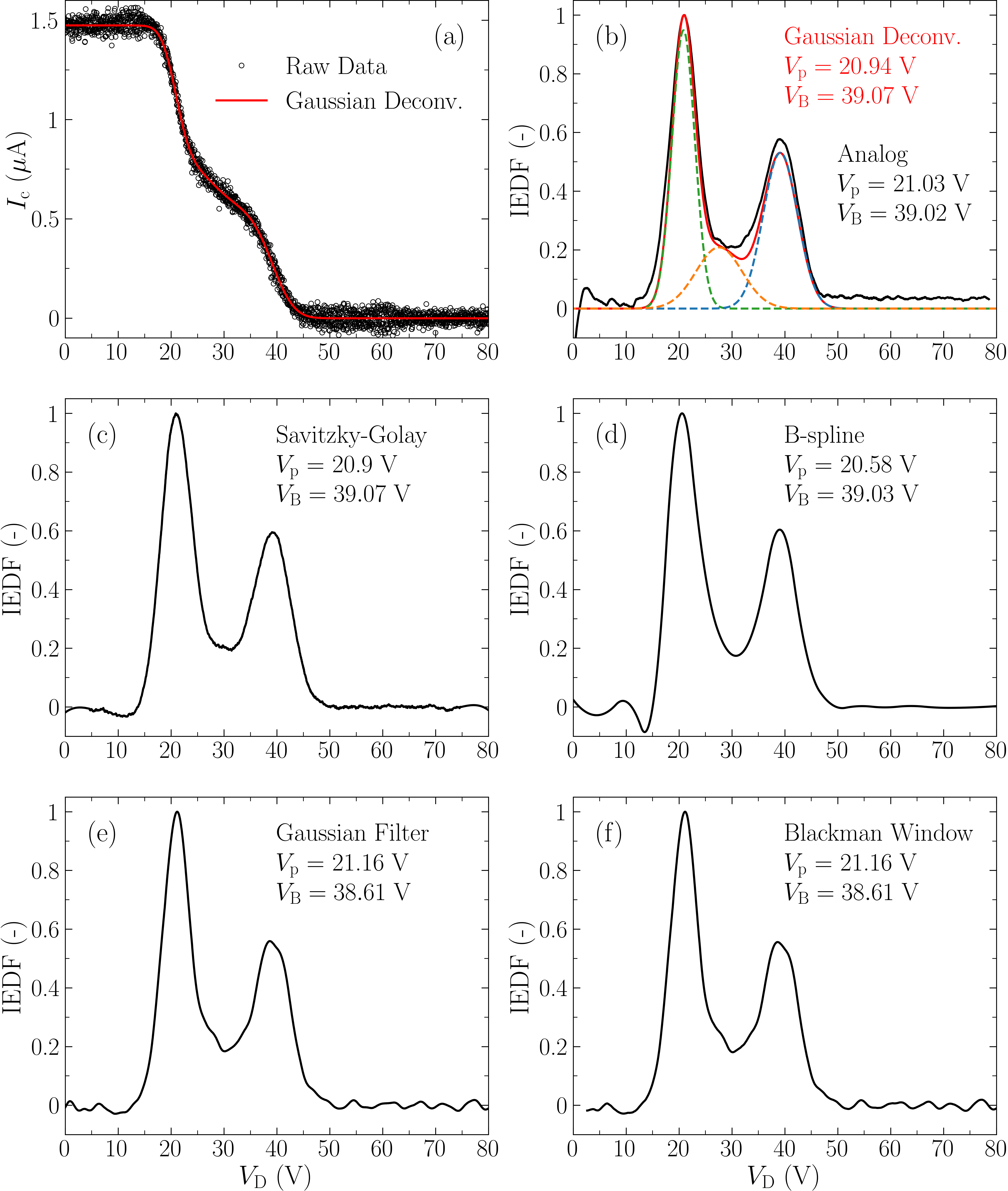}
\caption{Results of the analog and numerical differentiation and filtering methods from RFEA measurements showing the raw $I$-$V$ characteristics with the integrated Gaussian deconvolution curve (a), and the ion-energy distribution functions against the discriminator voltage obtained with: the analog differentiator (black line) and the Gaussian deconvolution method (red line) (b), the Savitzky-Golay filter (c), The B-spline polynomial (d), the Gaussian filter (e), and the Blackman Window filter (f). The dashed curves in (b) represent the three fitted curves resulting from the Gaussian deconvolution method.}
\label{fig:IEDF}
\end{figure*}

\Tabref{IEDF_EEFF_FilterParameters} summarizes the results of the parametric analysis that provide the smoothest IEDF possible for the numerical smoothing methods analyzed, while keeping intact the peak features. \figureref{IEDF} shows the ion-energy distribution functions obtained with the different analog and numerical techniques from experimental data. The IEDFs plotted are normalized by their peak value, and the position of the two peaks, i.e. the local plasma potential and the ion beam potential, is also reported. The data in \figref{IEDF} has a signal-to-noise ratio of approximately $66~$dB. The raw $I$-$V$ curve, shown in \figref[a]{IEDF}, was obtained by averaging 200 consecutive probe sweeps and it is plotted with the fitted Gaussian deconvolution to illustrate one of the data processing methods. 
As seen in \figref{IEDF}, the ion-energy distribution functions show a double-peak feature indicating the presence of a bi-population of ions: the first peak represents the background ion population (at zero energy) which location can be considered the local plasma potential $V_{\rm p}$, while the second peak corresponds to the accelerated ion beam population with potential $V_{\rm B}$ end energy $\varepsilon_{\rm B} = e (V_{\rm B} - V_{\rm p})$. Comparing measurements taken with a source-facing and a radial-facing RFEA shows that the IEDF does not present a peak separation effect due to RF modulation, but it effectively detects the presence of a directional ion beam \cite{Caldarelli2022JEP}. It is also noted that the ion-neutral mean free path for the operating pressure of $0.5~$mTorr is shorter than the distance between the source exit and the probe position i.e., $\lambda_{\rm i} \sim 10$\,cm and $z_{\rm p} = 19$\,cm. Thus, the accelerated ion beam component is decreased at the measurement location while the local, background ion population is enhanced due to charge-exchange collisions. Furthermore, elastic collisions could affect the beam direction, further reducing the collected beam current.

\Tabref{IEDF_perf} presents the key results obtained with the different data processing techniques, namely: the local plasma potential $V_{\rm p}$, the ion beam potential $V_{\rm B}$, the mean squared error (MSE), and the Shapiro-Wilk test statistic $W$. The Gaussian deconvolution method required the fitting of three Gaussian curves to accurately model the two ion populations. 

\begin{table}[H]
\renewcommand{\arraystretch}{1.0}
\caption{Results of the different IEDF processing methods in \figref[a]{IEDF}.}
\label{tab:IEDF_perf}
\centering
\setlength{\tabcolsep}{0.25em} 	
\begin{tabular}{l c c c c} \toprule	
       Method  &  $V\textsubscript{p}$\,(V) & $V\textsubscript{B}$\,(V) &  MSE$ / 10^{-4}~\rm(\mu A^{2})$ & $W$\\
       \midrule
        Analog & 21.03 & 39.02 & - & -\\
        Gaussian deconv. & 20.94 & 39.07 & $9.16$ & 0.9990\\
        SG & 20.90 & 39.07 & $8.97$ & 0.9986\\
        B-spline & 20.58 & 39.03 & $9.25$ & 0.9989\\
        GF & 21.16 & 38.61 & $9.08$ & 0.9984\\
        BW & 21.16 & 38.61 & $9.15$ & 0.9984\\
     \bottomrule
\end{tabular}
\end{table}

\begin{table*}[t]
\renewcommand{\arraystretch}{1.0}
\caption{Set of parameters for the processing used of the $I$-$V$ characteristics in \figref{IEDF} and \figref{EEPF}.}
\label{tab:IEDF_EEFF_FilterParameters}
\centering
\setlength{\tabcolsep}{0.3em} 	
\begin{tabular}{l c c c c c c c c c} \toprule	
    \multirow{2}{*}{Method} & 
    \multirow{2}{*}{Differentiation} & 
    \multicolumn{2}{c}{Window size $M$} & \multicolumn{2}{c}{Degree $N$} & \multicolumn{2}{c}{Sigma $\sigma$}  & \multicolumn{2}{c}{Knots $k$}\\ \cmidrule(lr){3-10}
    & & EEPF & IEDF & EEPF & IEDF& EEPF & IEDF& EEPF & IEDF\\ \cmidrule(lr){1-10}
        Savitzky-Golay Filter & Analytic & $115$ & $301$ & $2$ & $4$ & - & - & - & -\\
        B-spline Fitting & Analytic & - & - & $5$ & $3$ & - & - & $42$ & $15$ \\
        Gaussian Filter & Analytic & - & - & - & - & $25$ & $24$ & - & -\\ 
        Blackman Window & Central Difference & $156$ &  $140$ & - & - & - & - & - & -\\
    \bottomrule
\end{tabular}
\end{table*}

Comparing the analog differentiation method and the numerical techniques, it is evident that both the plasma and ion beam potentials are approximately unchanged with a standard deviations for $V_{\rm p}$ and $V_{\rm B}$ of $\pm0.21\,$V and $\pm0.23\,$V, respectively.
For the collected data, all the smoothing methods perform well in terms of mean squared error and the SW test statistic. The MSEs are $\leq 9.25 \times 10^{-4}~\mu A^2$ for all techniques, while the calculated $W$ values are all higher than 0.998. The Saviztky-Golay filter yields the lowest MSE, while the Gaussian deconvolution methods performs the best in the SW test, resulting in the highest $W$. It has to be noted that the Blackman filter showed an overshoot of the fitted polynomial at the edges caused by the zero padding in the convolution; this was addressed by disregarding the initial data points as the plasma characteristics in that region were not of interest.

\subsection{Electron-Energy Probability Function (EEPF)}

The RF compensated LP data were acquired with the apparatus operated at $200~$W of RF power, $1~$mTorr of argon, $B_{z,\rm max} = 300~$G and with the Helmholtz solenoids placed 30$~$cm away from the loop antenna. These conditions create an inductively coupled plasma exhibiting a symmetrical single peaked axial plasma density gradient ranging from $10^{12}~\rm cm^{-3}$ under the solenoids to $10^{10}~\rm cm^{-3}$ at the extremities of the glass tube \cite{filleul2021characterization,filleul2022ion,bennet2019non}. The LP was placed on axis halfway between the antenna and the solenoids. The processing of $I_{\rm p}(V_{\rm bias})$ to obtain the EEPF with the analog and numerical methods was conducted following the steps described in \Secref{Stats}. \Tabref{IEDF_EEFF_FilterParameters} gives the resulting optimal numerical parameters.

\figureref[a]{EEPF} shows the raw $I$-$V$ LP characteristic together with the optimized Savitkzy-Golay filter output from \figref{SG_NumParamVar}, i.e. for $(N=2,M=115)$. As reference, the signal-to-noise ratio of the LP characteristic is $48~$dB. In \figref[b)-(f]{EEPF}, the EEPFs obtained from the various methods and the deduced plasma parameters are shown. $T_{\rm eff}$ and $n_{\rm e}$ are calculated by integrating the EEPF according to \Eqref{Teff} and \Eqref{ne}, respectively. Their values are reported together with $V_{\rm p}$, the MSE, $W$, dynamic range and energy resolution for each method in \Tabref{EEPF_Perf}. The maximum variance in the obtained values of $V_{\rm p}$ is less than 2$\%$. Likewise, for $T_{\rm eff}$ and $n_{\rm e}$, their maximum variances are $\sim 13\%$ and $\sim 8\%$, respectively. Thus, in terms of estimating the plasma parameters, all 5 methods have comparable performance, i.e. within the absolute error tolerances usually associated with LP measurements \cite{boswell2015non,godyak2021rf}.

\begin{figure*}[!t]
\centering
\includegraphics[width=14.cm]{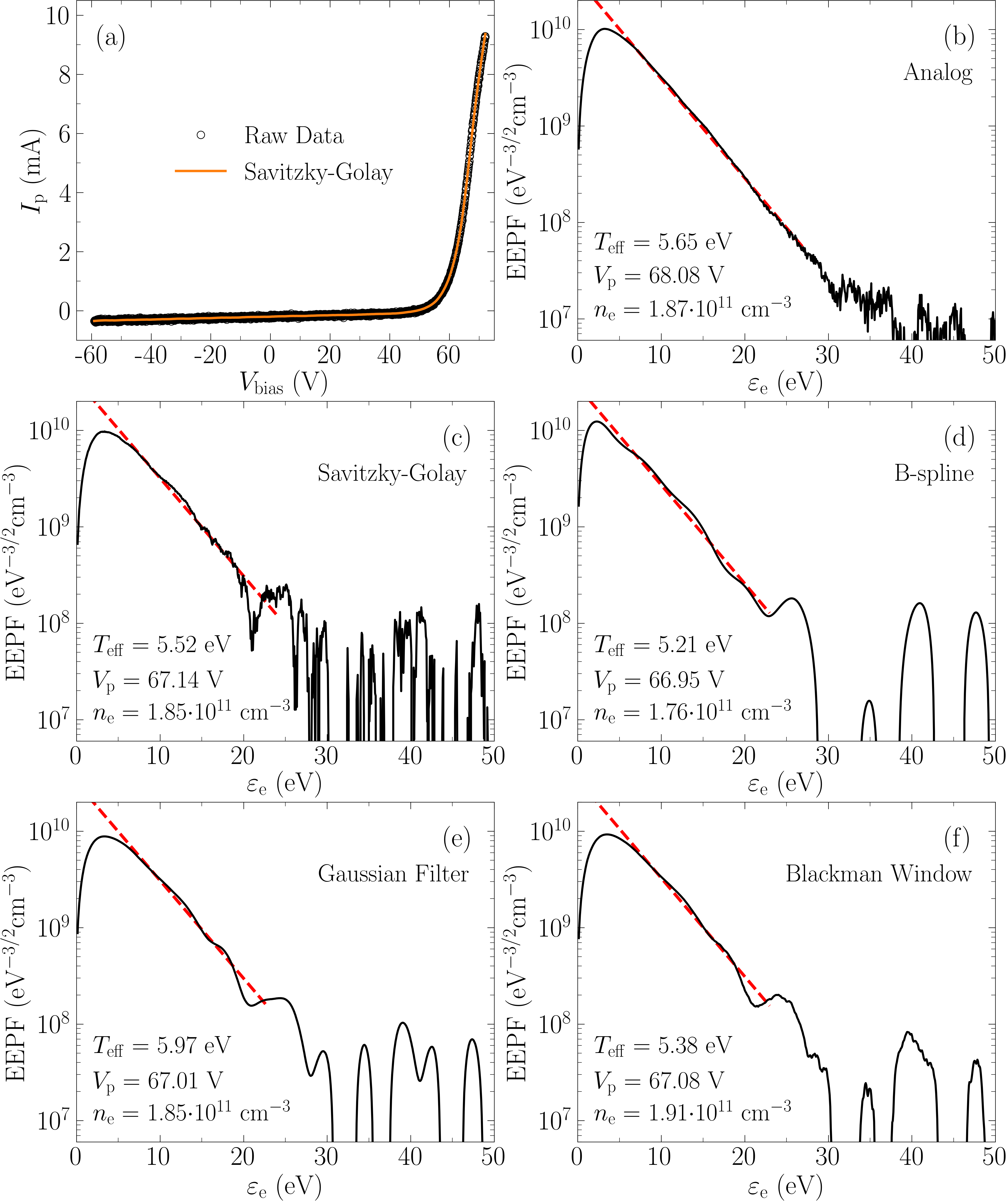}
\caption{Comparison of the analog and numerical methods for obtaining the EEPF, showing the raw $I$-$V$ characteristics (a), the EEPF obtained from the analog differentiator (b) and from the different numerical methods (c)-(f). The red dashed lines are the linear fit used to extract the electron temperature $T_{\rm e,fit}$.}
\label{fig:EEPF}
\end{figure*}

\begin{table*}[!t]
\renewcommand{\arraystretch}{1.0}
\caption{Results of the different EEPF processing methods in \figref{EEPF}.}
\label{tab:EEPF_Perf}
\centering
\setlength{\tabcolsep}{0.3em} 	
\begin{tabular}{l c c c c c c c} \toprule	
       Method  &  $V_{\rm p}~$(V) & $T_{\rm eff}~$(eV) & $n_{\rm e} / 10^{11}~\rm (cm^{-3})$ & MSE$ / 10^{-3}~\rm(mA^{2})$ & $W$ & DR$~$(dB) & $\delta_{\varepsilon_{\rm e}}~$(eV) \\
       \midrule
        Analog & 68.08 & 5.65 & 1.87 & - & - & 55 & 3.2 \\
        SG & 67.14 & 5.52 & 1.85 & 0.91 & 0.9964 & 32 & 3.4 \\
        B-spline & 66.95 & 5.21 & 1.76 & 0.84 & 0.997 & 36 & 2.3 \\    
        GF & 67.01 & 5.97 & 1.85 & 3.3 & 0.8015 & 33 & 3.4\\
        BW & 67.08 & 5.38 & 1.91 & 3.1 & 0.8537 & 33 & 3.5 \\
     \bottomrule
\end{tabular}
\end{table*}

In terms of energy resolution and dynamic range, different methods perform best depending on whether the focus is on resolving low-energy or high-energy electrons dynamics and properties. The B-spline fitting in \figref[d]{EEPF} resulted in the best energy resolution at $\delta_{\varepsilon_{\rm e}}\simeq 2.3~$eV and is the only method to satisfy the $\delta_{\varepsilon_{\rm e}} \leq (0.3-0.5)T_{\rm e}$ condition. The second-best performing method is the analog differentiation with $\delta_{\varepsilon_{\rm e}}\simeq 3.2~$eV, i.e. 0.375$~$eV away from satisfying the energy resolution condition. Regardless of the employed method, the energy resolution is ultimately limited by the absence of internal resistance compensation in the present LP implementation, in particular of the plasma-wall sheath impedance, or by the probe tip surface condition \cite{godyak2021rf}. Techniques to further improve the resolution are available in Ref.\,\onlinecite{ROH20151173,godyak2021rf}.

The analog differentiation method in \figref[b]{EEPF} resulted in an EEPF with a significantly larger dynamic range of approximately 3 orders of magnitude, compared to 2 orders of magnitude for the numerical methods. The analog differentiation is therefore the only method presented which can reliably resolve inelastic processes and can provide confidence on the nature of the electron distribution function, i.e. a single-Maxwellian for $\varepsilon_{\rm e}$ up to 30$~$eV in the present case. This is further highlighted in \figref[b]{EEPF} by the linear regression of the EEPF, shown by a dashed red line. The same linear regression applied to the numerically derived EEPFs highlights that the reduced DR and artificial oscillations in the case of the B-spline would make it difficult to reach conclusions on the nature of the electron distribution function. Among the numerical methods, the Savitzky-Golay and the B-spline are the best performing for this data set, with the lowest distortion of the original $I$-$V$ curve, as shown by their lower mean-square errors and Shapiro-Wilk test statistics in \Tabref{EEPF_Perf}. The Gaussian and Blackman window filters both resulted in noticeable distortions of the data ($W < 0.996$, i.e. their residuals are not normally distributed), in particular around the plasma floating potential and plasma potential portions of the $I$-$V$ curve.

Finally, it is interesting to note that following the guidelines in Ref.\,\onlinecite{godyak2021rf}, the low-energy gap $\delta_{\varepsilon_{\rm e}}$ impacts can be corrected by extrapolating the EEPF from its linear portion down to $\varepsilon_{\rm e}=0~$eV when the right conditions are satisfied. Because electron-electron collisions are the dominant processes for thermalizing electrons and making the distribution Maxwellian, and since the frequency of these collisions is inversely proportional to $\varepsilon_{\rm e}^{3/2}$, if the EEPF is Maxwellian for some $\varepsilon_{\rm e}$ greater than the energy of EEPF peak, then the EEPF is likely to be Maxwellian at lower energies \cite{tsendin2009electron,kaganovich2009non,godyak2021rf}. This extrapolation can be achieved with the linear regressions shown in \figref{EEPF} as dashed red curves.
Applying this process to the analog EEPF in \figref[b]{EEPF} for example, the effective electron temperature obtained with the corrected EEPF is $T_{\rm eff'}=4.32~$eV and the corrected electron density $n_{\rm e'}=2.57\times 10^{11}~\rm cm^{-3}$, to compare with the values in \Tabref{EEPF_Perf}. The electron temperature obtained from the slope of the linear regression is $T_{\rm e,fit}=4.2~$eV, providing evidence that the low-energy extrapolation has closely recovered the original EEPF shape.

\section{Conclusions}\label{sec:Conc}

This review outlines some of the most frequently employed data processing methods to obtain ion and electron-energy distribution functions from $I$-$V$ characteristics measured with a retarding field energy analyzer and an RF compensated Langmuir probe, respectively. Design recommendations relevant to measurements in RF magnetized plasmas are presented for both instruments. After describing the different processing techniques, an example of a parametric analysis to obtain optimal IEDFs and EEPFs is presented and applied to experimental data sets collected in a magnetized RF plasma apparatus. The recommended parameters optimization involves checking the mean-squared error and the normal distribution of the residuals (i.e. the Shapiro-Wilk test), coupled with a visual assessment process to ensure sufficient noise reduction with minimal distortion of the curves.

With respect to the IEDF, the analysis focused on the evaluation of first derivatives that accurately represented a plasma containing a bi-population of ions. For the data set under study, both the analog and numerical methods provide smooth IEDFs and consistent values of the local plasma and the ion beam potentials. Regarding the numerical methods analyzed, they all perform well in terms of MSE and $W$; in particular, the Saviztky-Golay filter and the Gaussian deconvolution methods delivered the best results in terms of obtaining accurate IEDFs with the least data distortion.

The EEPFs obtained from each method are assessed in terms of maximizing the energy resolution and dynamic range while minimizing distortions of the Langmuir probe $I$-$V$ curve. Among the numerical methods, the Savitzky-Golay filter and the B-spline fitting performed best in terms of distortion and energy resolution. All numerical methods delivered a mediocre dynamic range of approximately 2 orders of magnitude, which is to be related to the experimental data set employed having a low signal-to-noise ratio of $48~$dB. This could make it challenging to confidently identify the nature of the electron distribution function and to study inelastic electron processes. This is in contrast with the EEPF obtained from the same $I$-$V$ curve using the analog differentiator which showed a dynamic range of approximately 3 orders of magnitude and can be clearly identified as being Maxwellian up to $\sim 30~$eV. When the intent of calculating the EEPF is solely to obtain the plasma parameters, all methods explored are found to deliver equivalent results. The analog differentiator technique provides an appreciable reduction in the IEDF and EEPF processing time compared to the numerical methods since no data-set specific optimization is required.

\section*{Data Availability Statement}
The data that support the findings of this study are available from the corresponding author upon reasonable request.

\section*{Author Contributions}
In alphabetical order, A.C. and F.F. share first authorship. A.C. and F.F. performed the data acquisition, data processing and manuscript preparation concerning the IEDF and EEDF, respectively. R.B. and C.C. provided guidance in experimental RF plasmas and diagnostics. N.R. and J.C. are doctoral supervisors of A.C. and F.F.

\section*{Conflict of Interest Statement}
The authors declare that the research was conducted in the absence of any commercial or financial relationships that could be construed as a potential conflict of interest.

\bibliography{./References}


\end{document}